\documentclass[11pt,twoside]{book}
\usepackage{vshalo}
\usepackage{longtable}
\usepackage{amsmath,amssymb}
\usepackage{graphicx}
\usepackage{lscape}
\usepackage{epsf}
\usepackage{index}
\usepackage{natbib}
\usepackage{bigdelim}
\usepackage{multirow}
\makeindex
\newindex{aut}{adx}{and}{\Large Author  Index}
\newcommand{\axindex}[1]{\index[aut]{#1}}

\begin{document}

\pagestyle{myheadings}
\setcounter{equation}{0}\setcounter{figure}{0}\setcounter{footnote}{0}\setcounter{section}{0}\setcounter{table}{0}\setcounter{page}{35}
\markboth{Dambis}{Estimating the Kinematic Parameters and the Distance-Scale Zero Point}
\title{Estimating the Kinematic Parameters and the Distance-Scale Zero Point for the Thin-Disk, Thick-Disk, and Halo Population Tracers via 3D Velocity Data}
\author{A. K. Dambis}\axindex{Dambis, A. K.}
\affil{Sternberg Astronomical Institute, Universitetskii pr. 13, Moscow, 119992 Russia} 

\begin{abstract}We use the method of statistical parallax to constrain the distance-scale zero points and analyze
the kinematics of extensive samples of Galactic classical Cepheids, RR Lyrae type variables, and blue horizontal branch 
stars, which serve as standard 
candles/kinematic tracers of various Galactic populations. We obtain three consistent estimates for the local
circular velocity based on the mean velocities of halo RR Lyrae variables, BHB stars, and Galactic rotation curve 
inferred from Cepheid data  with an average value of 210~$\pm$~6~km~s$^{-1}$, which is close to the average circular 
velocity in the 5--40~kpc interval of Galactocentric
distances inferred from BHB star data (195~$\pm$~5~km~s$^{-1}$), thereby providing further supporting evidence 
for the practically flat shape of the Galactic rotation curve beyond $\sim$~5~kpc from the center. The inferred
distance-scale corrections imply a solar Galactocentric distance of 7.7~$\pm$~0.4~kpc, an LMC distance modulus of 
18.42~$\pm$~0.06, and a Hubble constant of 73--85~km~s$^{-1}$~Mpc$^{-1}$.

\end{abstract}

\section{Introduction}
Establishing the kinematics of the major Galactic subsystems --- the halo and the thin and thick disk --- is a task
of fundamental importance,  bringing us closer to understanding the origin, evolution, and mass distribution in the 
Galaxy. In this work,
we choose Cepheids as our kinematic tracers for the young (thin) Galactic disk and RR Lyrae variables as the kinematic
tracers for the thick disk and halo. We further expand our halo kinematic-tracer base by including an extensive sample 
of blue horizontal-branch stars (BHB) spanning heliocentric (and Galactocentric) distances from $\sim$~5 to several tens
of kpc. 

Here we use the method of statistical parallaxes in its rigorous maximum-likelihood version suggested by 
\citet{Murray1983} (pp. 297--302) to estimate the kinematical parameters of the 
sample of Galactic Cepheids, RR Lyrae type variables, and BHB stars and to calibrate the underlying photometric
distance scales (based on the period-K-band luminosity relations for Cepheids and RR Lyraes and on the BHB star
distance scale proposed by \citet{Sirko_et_al_2004}, which is based on SDSS colors and the appropriate model
atmospheres). 

\section{Cepheids}
Our sample consists of 215 Cepheids with bona-fide proper motions from the new reduction of Hipparcos data 
\citep{van_Leeuwen2007}, accurate $\gamma$-velocities from Gorynya et al. (1992, 1996, 1998)
and photometry from Berdnikov~(1995, 1998) and Berdnikov et al.~(2000). The distances are based on the
Cepheid $K$-band PL relation derived by \citet{Berdnikov1996} and the procedure proposed in the latter paper. The 
results are listed in Table~1. Here $U_0$, $V_0$, and $W_0$ are the components of the local mean solar velocity toward
the Galactic center, in the direction of Galactic rotation, and toward the North Galactic Pole, respectively, 
relative to the sample considered, $\sigma V_R$, $\sigma V_{\phi}$, and $\sigma V_Z$ are the diagonal
components of the velocity dispersion tensor in the directions of projected Galactocentric radius, Galactic rotation,
and toward the North Galactic Pole, respectively. The parameters $\Omega_0$, A, and  $\Omega^{\prime \prime}$ are
equal to the local angular circular rotation velocity, Oort's constant A, and the second derivative of the local
angular circular rotation velocity with respect to projected Galactocentric distance, respectively, and $k$ is the 
distance-scale correction factor (to be multiplied by the initial distances to obtain the corrected distances).

\begin{table}[!ht] 
\caption{Kinematical and distance-scale solution for Cepheids (N=215 stars).
Here $U_0$, $V_0$, $W_0$, $\sigma V_R$, $\sigma V_{\phi}$, and $\sigma V_Z$ are in km~s$^{-1}$,
$\Omega_0$ and A are in km~s$^{-1}$~kpc$^{-1}$, and $\Omega^{\prime \prime}$ is in km~s$^{-1}$~kpc$^{-3}$.}
\begin{center}
\begin{tabular}{rrrrrrrrrr}
\tableline
\noalign {\smallskip} 
$U_0$ & $V_0$ & $W_0$ & $\sigma V_R$ & $\sigma V_{\phi}$ &
$\sigma V_Z$  & $\Omega_0$ & A & $\Omega^{\prime \prime}$ & k\\ 
\noalign{\smallskip}
\tableline
-8.5 & -12.2 & -7.1 & 13.0 & 9.3 & 8.9 & 27.5 & 15.7 & 0.80 & 1.18 \\
$\pm$~1.0 &$\pm$~0.9 & $\pm$~0.9 & $\pm$~0.8 &$\pm$~0.6 & $\pm$~0.9 & $\pm$~0.8 & $\pm$~0.7 &$\pm$~0.13  & $\pm$~0.06\\ 
\noalign{\smallskip}
\tableline
\noalign{\smallskip}
\end{tabular}  \end{center}
\scriptsize

\end{table}

\section{RR Lyrae type variables}

Our sample consists of 364 RR Lyrae type variables with bona-fide proper motions , radial velocities , 
and phase-corrected 2MASS K-band magnitudes. The distances are based on the
$K$-band PL relation derived by \citet{j1}.
See \citet{Dambis2009} for a complete description of the data set, which is available from
the CDS. We applied to it a bimodal version of the statistical-parallax method \citep{Dambis2009} and obtained the 
results that we list in Table~2.

\begin{table}[!ht] 
\caption{Kinematical solution and distance-scale correction for Galactic field RR Lyrae variables based on the
bimodal solution (364 stars with $r_{hel} <$ 6.5 kpc). Here $U_0$, $V_0$, and $W_0$, 
have the same meaning as in Table~1, and $\sigma V_R$, $\sigma V_{\phi}$, 
and $\sigma V_{\theta}$ are the diagonal components of the velocity dispersion tensor aligned with the spherical 
coordinate system centered on the Galactic center. All these quantities are in km~s$^{-1}$}.
\begin{center}
\begin{tabular}{l r  r r r r r r r }
\tableline
\noalign {\smallskip} 
Population & Fraction &  $U_0$ & $V_0$ & $W_0$ & $\sigma V_R$ & $\sigma V_{\phi}$ &
$\sigma V_{\theta}$  & k\\ 
\noalign{\smallskip}
\tableline
Halo & 0.75 & -12 & -217 & -6 & 167 & 86 & 78 &  \\
& 0.03  & $\pm$~10 &$\pm$~9 & $\pm$~6 & $\pm$~9 &$\pm$~6 & $\pm$~5 & 0.97\\ 
\noalign{\smallskip}
Thick disk & 0.25 & -15 & -45 & -25 & 55 & 44 & 30 & $\pm$~0.04 \\
& 0.03  & $\pm$~7 &$\pm$~7 & $\pm$~5 & $\pm$~6 &$\pm$~6 & $\pm$~4 &  \\ 
\tableline
\noalign{\smallskip}
\end{tabular}  \end{center}
\scriptsize
\end{table}

\section{BHB stars}

Our third sample consists of 1955 rigorously selected blue horizontal-branch halo stars from SDSS DR6~\citep{SDSS6} with
photometric distances based on the calibration of \citet{Sirko_et_al_2004}, radial velocities determined
within the framework of SEGUE project~\citep{SEGUE}. We adopted this sample from \citet{Xue_et_al2008}, and supplemented
it with absolute proper motions from the SDSS DR7~\citep{SDSS7} database. Here we use only stars sample 
with $R_g \leq$~40~kpc
and removed a number of evident outliers. 
We applied to the resulting sample a single-mode version of the statistical-parallax method 
parametrizing the variation of the velocity-ellipsoid parameters as a function of Galactocentric distance
in terms of the model of \citet{Sommer-Larsen1997} and obtained the results that we list in Table~3.

\begin{table}[!ht] 
\caption{Kinematical solution and distance-scale correction for the sample of BHB stars 
(1955 stars with $r_{hel} <$ 40 kpc). Here $U_0$, $V_0$, and $W_0$ have the same meaning as in Table~1 and
$\sigma_0$, $\sigma^+$, $l$, $r_0$, and $V_c$ are the parameters of the kinematical halo model of 
\citet{Sommer-Larsen1997}. The parameters $U_0$, $V_0$, $W_0$, $\sigma_0$, $\sigma^+$, and $V_c$ are in km~s$^{-1}$,
and $l$ and $r_0$
are in kpc. Note that $V_c$ is the circular rotation velocity of the (assumed) flat rotation curve and $r_0$ is 
the Galactocentric distance at which the shape of the velocity ellipsoid changes from radial to tangential anisotropy.}
\begin{center}
\begin{tabular}{ r r r r r r r r r }
\tableline
\noalign {\smallskip} 
 $U_0$ & $V_0$ & $W_0$ & $\sigma_0$ & $\sigma^+$ & $l$ &$r_0$ & $V_c$ & k\\ 
\noalign{\smallskip}
\tableline
-14 & -225 & -2 & 96 & 89 & 3.7 & 19.2 & 195 & 1.06\\
$\pm$~4 &$\pm$~5 & $\pm$~3 & $\pm$~12 &$\pm$~9 & $\pm$~0.4 & $\pm$~1.6& $\pm$~5 & $\pm$~0.03 \\ 
\tableline
\noalign{\smallskip}
\end{tabular}  \end{center}
\scriptsize
\end{table}

\section{Distance scale}
In Table~4 we list the distance-scale correction factors inferred via statistical parallax ($k$) and 
and the average distance-scale correction factors based on statistical and Hipparcos trigonometric
parallaxes ($k_2$) for Cepheids and RR Lyrae variables, as well as the corresponding LMC distance-modulus
estimates. We also give the estimate of the solar Galactocentric distance ($R_0$) based on the IR photometry 
of Galactic bulge RR Lyraes. The weighted average of the LMC distance modulus ($DM_{LMC}$=18.42$\pm$0.06)
implies a Hubble constant of 73--85~km~s$^{-1}$~Mpc$^{-1}$.

\begin{table}[!ht] 
\caption{Distance-scale results.}
\begin{center}
\begin{tabular}{lrrrr}
\tableline
\noalign {\smallskip} 
Distance indicator  & $k$   & $k_2$            & $DM_{LMC}$       & $R_0$, kpc\\ 
\noalign{\smallskip}
\tableline
Cepheids & 1.18~$\pm$~0.06  &  1.12~$\pm$~0.03 & 18.50~$\pm$~0.07 &      \\ 
\noalign{\smallskip}
RR Lyraes& 0.97~$\pm$~0.04  &  0.98~$\pm$~0.04 & 18.30~$\pm$~0.09 &  7.7~$\pm$~0.4  \\ 
\tableline
\end{tabular}  \end{center}
\end{table}

\section{Circular velocity}
Our kinematical analysis yields three independent estimates for the local velocity of circular Galactic rotation,
$V_c(local)$:  $V_c(local) =|V_0(halo)-V_0(Cepheids)|$, where $V_0(halo)$ can be 
represented either by halo RR Lyraes or BHB stars (the corresponding two $V_c(local)$ values are equal to 
205~$\pm$~9 and 205~$\pm$~13~km~s$^{-1}$, respectively), or as 
$V_c(local) = |R_0 \cdot \Omega_0|$~=~212~$\pm$~13~km~s$^{-1}$. They agree rather well with each other
and their weighted average $V_c(local)$~=~210~$\pm$~6~km~s$^{-1}$ differs little from the global value of
$V_c(global)$~=~195~$\pm$~5~km~s$^{-1}$ inferred from an analysis of the kinematics of BHB stars out to a Galactocentric
distance of $R_g~\sim$~40~kpc, thereby further reinforcing the case of the flat Galactic rotation curve beyond 
$R_g~\sim$~5~kpc.

\section{Acknowledgments}
This work was supported by the Russian Foundation for Basic Research (projects nos.~07-02-00380-a and 08-02-00738-a)
and the Council for the Program of Support for Leading Scientific Schools (project no.~NSh-433.2008.2).


\begin{thebibliography}{}  
\bibitem[Berdnikov(1995)]{Berdnikov2005} Berdnikov L.N., Cepheid Data Bank, ASP Conf. Ser., 1995, 349
\bibitem[Berdnikov(2008)]{Berdnikov2008} Berdnikov L.N., Photoelectric observations of Cepheids in UBVRI. Vizier 
online data.  Catalog CDS II/285, 2008.
\bibitem[Adelman-McCarthy~et~al.~(2008)]{SDSS6} Adelman-McCarthy~J.K.~et~al., 2008, \apjs, 175, 297
\bibitem[Abazajian~et~al.~(2009)]{SDSS7} Abazajian K.N.~et~al., 2008, \apjs, 182, 543
\bibitem[Berdnikov et al.~(1996)]{Berdnikov1996} Berdnikov L. N., Vozyakova O. V., Dambis A. K.,
Astron. Lett., 1996, 22, 838
\bibitem[Berdnikov et al.~(2000)]{Berdnikov2000} Berdnikov L. N., Dambis A. K., Vozyakova O. V., 2000, Astron.
Astrophys. Suppl., 143, 211
\bibitem[Dambis(2009)]{Dambis2009} Dambis A.K., 2009, Mon. Notices Roy. Astron. Soc.,  396, 553
\bibitem[Gorynya et al.~(1992)]{Gorynya1992} Gorynya N. A., Irsmambetova T. R.,
Rastorgouev A. S., Samus N. N., Sov. Astron. Lett., 1992, 18, 316
\bibitem[Gorynya et al.~(1996)]{Gorynya1996} Gorynya, N. A., Samus N. N., Rastorgouev A. S., Sachkov  M. E.,  
Astron. Lett., 1996, 22, 175
\bibitem[Gorynya et al.~(1998)]{Gorynya1998} Gorynya, N. A., Samus N. N., Sachkov  M. E., Rastorgouev A. S., 
Glushkova E.V., Antipin S.V., Astron. Lett., 1998, 24, 815
\bibitem[Jones et al.~(1992)]{j1} Jones R. V., Carney B. W., Storm J.,
Latham D., 1992, ApJ, 385, 646
\bibitem[Murray(1983)]{Murray1983} Murray C.A., 1983, {\it Vectorial Astrometry},
(Bristol: A.Hilger)
\bibitem[Sirko et al.~(2004)]{Sirko_et_al_2004} Sirko E., Goodman J., Knapp G.R., 
Brinkmann J., Ivezic Z., Knerr E.J., Schlegel D., Schneider D.P., York D. G ., 2004, \aj, 127, 899
\bibitem[Sommer-Larsen et al.~(1997)]{Sommer-Larsen1997} Sommer-Larsen J., Beers T. C., Flynn C., Wilhelm R.,
Christensen P. R., 1997, \apj, 481, 755
\bibitem[van Leeuwen~(2007)]{van_Leeuwen2007} van Leeuwen F., 2007,  Hipparcos, 
the New Reduction of the Raw Data, (New York: Springer)
\bibitem[Xue et al.~(2008)]{Xue_et_al2008} Xue X. X., Rix H. W., Zhao G., Re Fiorentin P., Naab T.
Steinmetz M., van den Bosch F. C., Beers T. C., Lee Y. S., Bell E. F., Rockosi C., Yanny B., Newberg H., Wilhelm, R.,
Kang X., Smith M. C., Schneider D. P., \aj, 684, 1143
\bibitem[Yanny~et~al.~(2009)]{SEGUE} Yanny B.~et~al., 2009, \aj, 137, 4377
\end{thebibliography}
\end{document}